# Electrical transport and magnetic properties of nanostructured La$_{0.67}$Ca$_{0.33}$MnO$_3$

Y. G. Zhao,[a),c)] W. Cai,[a)], X. S. Wu,[b)] X. P. Zhang,[a)] K. Wang,[a)] S. N. Gao[a)] and L. Lu[b)]

[a)]*Department of Physics, Tsinghua University, Beijing 100084, P. R. China*

[b)]*Institute of Physics, Chinese Academy of Sciences, Beijing 100080, P. R. China*

Nanostructured La$_{0.67}$Ca$_{0.33}$MnO$_3$ (NS-LCMO) was formed by pulsed-laser deposition on the surface of porous Al$_2$O$_3$. The resistance peak temperature ($T_p$) of the NS-LCMO increases with increasing average thickness of the films, while their Curie temperatures ($T_c$) remain unchanged. The coercive field of the samples increases with decreasing film thickness and its temperature dependence can be well described by $H_c(T) = H_c(0)[1-(T/T_B)^{1/2}]$. A large magnetoresistance and strong memory effect were observed for the NS-LCMO. The results are discussed in terms of the size effect, Coulomb blockade and magnetic tunneling effect. This work also demonstrates a new way to get nanostructured manganites.

The discovery of colossal magnetoresistance (CMR) in the hole-doped manganese perovskites $L_{1-x}A_xMnO_3$, where L and A are trivalent lanthanide and divalent alkaline earth ions, has attracted much interest because of their interesting physics and the potential for magnetoelectronic applications.[1] Currently, one of the main trends in device applications is the size reduction of devices to nanoscale. It is therefore important to study the structure and properties of the CMR materials at such a nanoscale. Gong et al obtained nano-dots on the surface of $La_{0.5}Ca_{0.5}MnO_3$ thin films by annealing the films at high temperatures, and the magnetoresistance (MR) of the films was enhanced.[2] There have also been some reports on the preparation and properties of bulk CMR material with ultra-fine grains by low-temperature sintering or mechanical alloying, and the samples thus obtained were found to exhibit size effect.[3-7] In these techniques, the low-temperature sintering and mechanical alloying may lead to some other effects,[5] such as the reduction in the oxygen content. Also, such techniques can only get ultra-fine grains with three-dimensional neighbor connections.

The nanoscale network on the surface of porous $Al_2O_3$ has been used as a template to grow nanostructured Fe and NiFe,[8,9] which show size effect. It is also interesting to study the growth of nanostructured CMR on such a template and the corresponding physical properties of the samples. In this paper, we report the growth of NS-LCMO on the surface of porous $Al_2O_3$ template by pulsed-laser deposition and its interesting electrical transport and magnetic properties.

$La_{0.67}Ca_{0.33}MnO_3$ (LCMO) films were deposited at 750 °C on porous $Al_2O_3$ substrates by pulsed-laser deposition using a KrF laser ($\lambda$=248 nm). The diameter of the pores is 0.2 μm. The deposition pressure, energy density and repetition rate were 40 Pa, 2 J/cm$^2$ and 5 Hz, respectively. After the deposition, the chamber was filled with 1 atm oxygen immediately. The sample was cooled down to 600 °C, and then kept at this temperature for 20 minutes, followed by a cooling down to room temperature. The morphology of the films was found to be controlled by the deposition time. The average thickness of the films was obtained by calibration on films grown on crystalline LaAlO$_3$. The phase analysis of the samples was performed using a Rigaku D/max-RB x-ray diffractometer with Cu kα radiation. The electrical resistivity was measured by the four probe method. A SQUID magnetometer was used to measure the temperature dependence of magnetization with the magnetic field parallel to the sample surface. MR was measured from 300 K to 4 K. The morphology of the films was studied using field emission SEM.

Fig.1 shows the surface morphology of the porous Al$_2$O$_3$ (a), NS-LCMO films with average thickness of 36 nm (b) and 180 nm (c), respectively. Porous Al$_2$O$_3$ shows the pores with diameter of about 200 nm and the nanoscale walls surrounding the pores. It can be seen that NS-LCMO was formed on the surface of porous Al$_2$O$_3$. The morphology of the NS-LCMO changes gradually with the average thickness, which is controlled by the deposition time. For 36 nm thick sample, LCMO nano rings were formed. The grain size of the samples decreases with decreasing thickness. It should be pointed out that the nano-grains here are connected in one or two dimension, which is quite different from the three dimensional connection in the bulk samples with ultra-fine grains.[3-7] X- ray diffraction pattern for NS-LCMO with average thickness of 180 nm is consistent with that of the LCMO bulk sample, indicating that the NS-LCMO is single phased.

Figure 2 shows the temperature dependence of resistance for the NS-LCMO films with different average thickness. With the increase of the average thickness of the films, $T_p$ increases while $T_c$ remains unchanged (not shown here). The upturn of the resistance at low temperatures was also observed in $La_{2/3}Sr_{1/3}MnO_3$ bulk samples with ultra-fine grains and was attributed to the Coulomb blockade.[3] The inset of Fig. 2 shows that the resistance at low temperatures does follow the $\exp(E_c/T)^{1/2}$ law, where $E_c$ is the charging energy, manifested in the Coulomb blockade mechanism. For the resistance peak, we argue that it is not related to the insulator-metal transition as believed.[3] Instead it is related to the temperature dependence of the magnetic tunneling effect between grains, which determines the peak resistance and peak temperture. The reason is that the resistivity of NS-LCMO is about 15 $\Omega\cdot$cm, which is four orders higher than the Ioffe-Regel limit (~2m$\Omega\cdot$cm) [ref. 10] for metallic oxide materials. Since with temperature decreasing, the alignment of the magnetic moments is expected to improve, the magnetic tunneling effect will make the resistance decrease. In contrast, the Coulomb blockade increases the resistance as temperature decreases. However, Coulomb blockade is only significant at low temperatures. The total resistance of NS-LCMO is proportional to the contribution of magnetic tunneling times the contribution of electronic tunneling (here is the Coulomb blockade).[11] These contributions result in the resistance peak and the resistance upturn at low temperatures. In $La_{0.85}Sr_{0.15}MnO_3$ bulk samples, it has been shown that the magnetic tunneling between grains can result in the resistance peak, which is not due to the insulator-metal transition.[12]

Figure 3 shows the temperature dependence of magnetoresistance for 36 nm thick film and the inset is its magnetization in zero field cooling (ZFC) and field cooling (FC) status, respectively. From 5 K to about 100 K, the MR, defined as [R(0)-R(H)]/R(0), is 100%. Above 100 K, MR decreases linearly with increasing temperature. The MR of NS-LCMO can be explained in terms of the decrease of the tunneling barrier height induced by external magnetic field, which decreases the resistance of NS-LCMO. From the inset of Fig.3, it can be seen that the ZFC and FC curves show some dramatic difference and the paramagnetic-ferromagnetic transition is not sharp. These behaviors have been seen in the ferromagnetic nano-particles.[13] The difference between the ZFC and FC curves is due to the blocking of the magnetic moment of the nano-particles below a characteristic blocking temperature ($T_B$). Between $T_c$ and $T_B$, the sample is in the superparamagnetic regime.

Figure 4 shows the temperature dependence of the resistance with cooling and warming for 36 nm thick LCMO film. The sample was first cooled down to obtain the cooling R-T curve. Then a 14 Tesla magnetic field was applied at 5 K for 2 hours. Afterwards, the magnetic field was removed and the sample temperature was increased to obtain the warming R-T curve. The temperature dependence of the resistance with cooling and warming for 180 nm thick LCMO film is shown in the inset, which was measured in the same way as the 36 nm thick LCMO film. For 36 nm thick LCMO film, the cooling and warming R-T curves show dramatic difference, indicating that the magnetic field strongly reduced the resistance of the sample. This change cannot be recovered after the removal of the magnetic field, showing a strong memory effect. It is also noted that $T_p$ shifts to 100 K in the warming curve with a 40 K increase compared with the cooling curve. In contrast, for thick LCMO film, the cooling and warming R-T curves at low temperatures only show a small difference. So the magnetic field memory effect for the thicker sample is not remarkable.

Magnetic field memory effect has been observed in charge ordered (CO) manganites[14, 15] and $Nd_{0.7}Sr_{0.3}MnO_3$[16], where the application of a magnetic field suppresses the barrier height between the two bi-stable states of the materials, made it possible to switch from the charge "crystal" state (the CO state) to the charge "liquid" state. Upon removal of the field, the materials will not recover to their original state due to the high barrier height then. For our case, although there is no CO phase involved, the underline mechanism of the memory effect could be similar. It may arise from the field alignment of the magnetic domains or moments in a frustrated local environment in our sample, due to the grain size effect and the particular way of connection between the grains. It has been shown that a surface spin glass layer could exist for grains in NS-LCMO.[17] Existence of surface spin glass has been proved in $La_{2/3}Sr_{1/3}MnO_3$ nanopaticles.[18]

Shown in Fig. 5 is the temperature dependence of the coercivity ($H_c$) for 36 nm and 180 nm thick LCMO films. $H_c$ of the 36 nm thick LCMO film is larger than that of the 180 nm thick film below $T_c$. As seen in Fig.1, the average grain size of the 36 nm thick LCMO film is smaller than that of the 180 nm thick film. This grain size dependence of $H_c$ is consistent with previous reports on soft ferromagnetic materials,[19] which shows that when the average grain size (d) of the material is larger than its critical grain size ($d_c$), $H_c$ follows the 1/d-law, which is related to an increase in the magneto-crystalline anisotropy. The temperature dependence of $H_c$ for the two samples can be well described by[20] $H_c(T) = H_c(0)[1-(T/T_B)^{1/2}]$, a relation that has been well established in nano feromagnetic metals and alloys in the superparamagnetic regime and below the blocking temperature $T_B$. To our knowledge, this relation has not been reported before in manganites. It is interesting to note that in NS-LCMO, $T_B$ is very close to $T_c$, which would indicate that the superparamagnetic domains or clusters are easily blocked by their surroundings, preventing them from being free rotating. This is in agreement with the core-shell picture for our samples where the ferromagnetic core is surrounded by spin glass shells [ref. 17].

In summary, nanostructured LCMO samples were prepared with better controlled grain size and the connections between the grains. The NS-LCMO shows some peculiar electrical transport and magnetic properties, including the strong external magnetic field memory effect, temperature dependence of the coercive field and huge MR, etc. The present work also provides a new way to get nanostructured manganites.

This work was supported by NSFC (project No. 50272031), the Excellent Young Teacher Program of MOE, P.R.C, 973 project (No. 2002CB613505), Specialized Research Fund for the Doctoral Program of Higher Education (No. 2003 0003088) and the Knowledge Innovation Program of the Chinese Academy of Sciences. YGZ thanks Profs. T. Venkatesan, S. Ogale, X. X. Xi and D. S. Dai for enlightening discussion.

**Figure Captions**

Fig.1 Surface morphology of the porous $Al_2O_3$ (a) and NS-LCMO films with average thickness of 36 nm (b) and 180 nm (c), respectively.

Fig.2 Temperature dependence of resistance for the NS-LCMO films with different thickness. The inset gives the $T^{-1/2}$ dependence of logR.

Fig.3 Temperature dependence of magnetoresistance for 36 nm thick film. The inset shows the magnetization with zero field cooling (ZFC) and field cooling (FC), respectively.

Fig.4 Temperature dependence of resistance with cooling and warming for 36 nm thick sample and 180 nm thick sample (inset).

Fig.5 Variation of $H_c$ with temperature for 36 nm and 180 nm thick samples. The inset shows the linear dependence of $H_c$ on $T^{1/2}$.

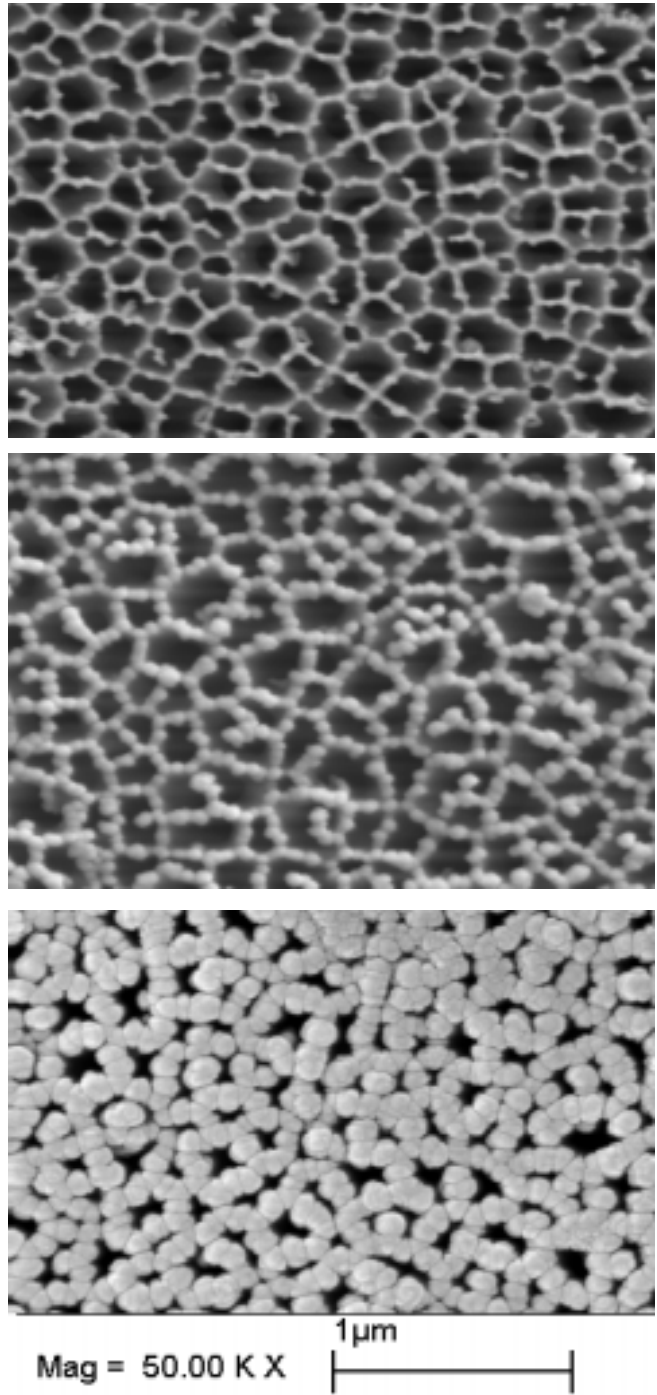

Fig.1

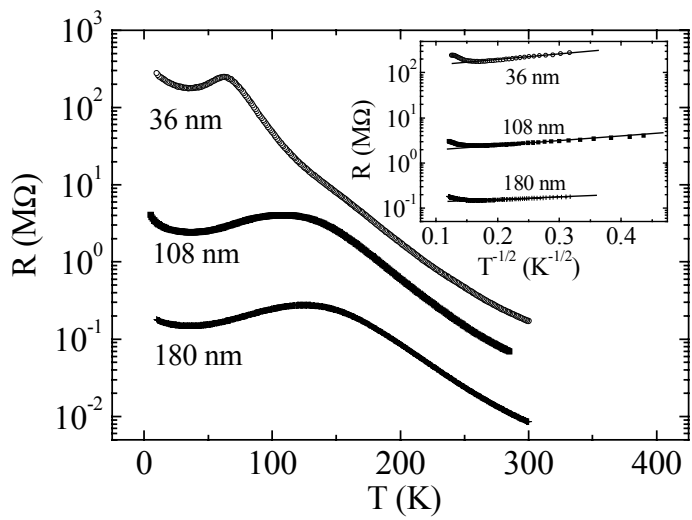

Fig.2

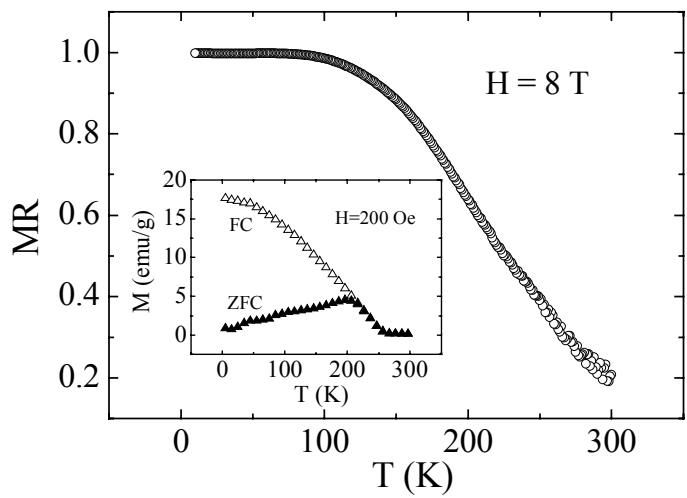

Fig.3

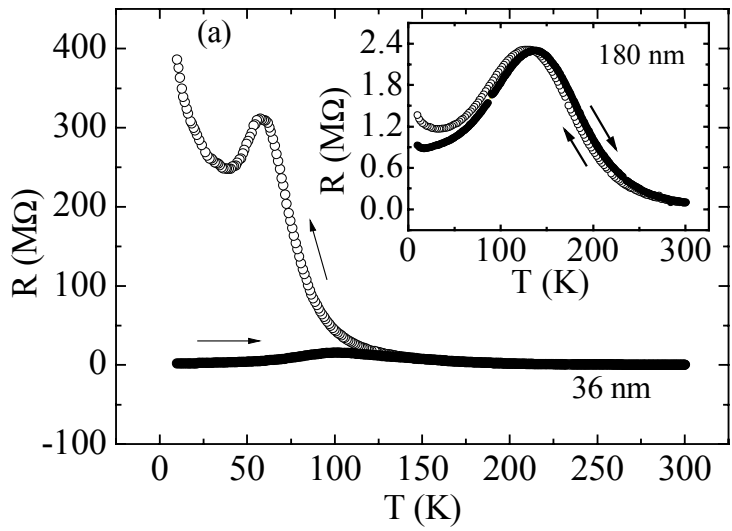

Fig.4

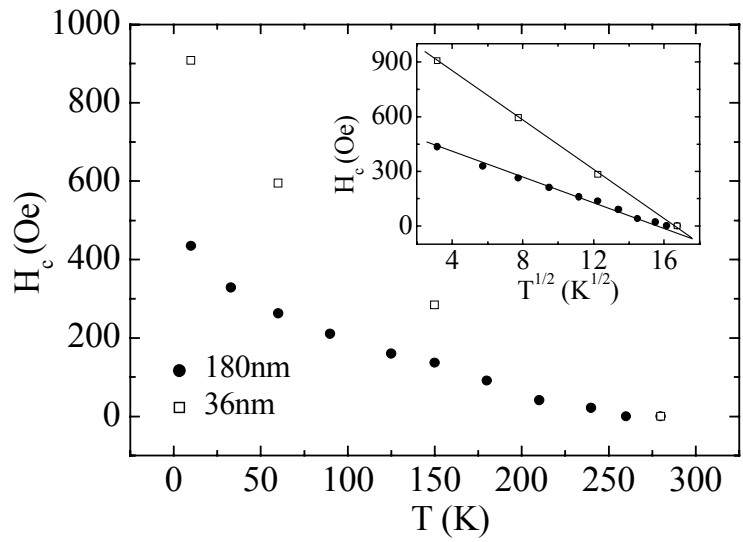

Fig.5